\documentclass[conference]{IEEEtran}
\IEEEoverridecommandlockouts

\usepackage[utf8]{inputenc}
\usepackage[T1]{fontenc}
\usepackage[american]{babel}
\usepackage[center]{caption}
\usepackage{graphicx}
\usepackage{listings}
\usepackage{float}
\usepackage{amssymb,amsmath,amsfonts,amsthm,bm,exscale}
\usepackage{blindtext, graphicx}
\usepackage{verbatim}
\usepackage{algorithm}
\usepackage{algpseudocode}
\usepackage{fancyvrb}
\usepackage{bera}
\usepackage{amsmath}
\usepackage{mathtools}
\usepackage{lipsum}
\usepackage{scalerel}
\usepackage{tikz}
\usetikzlibrary{svg.path}

\definecolor{orcidlogocol}{HTML}{A6CE39}
\tikzset{
  orcidlogo/.pic={
    \fill[orcidlogocol] svg{M256,128c0,70.7-57.3,128-128,128C57.3,256,0,198.7,0,128C0,57.3,57.3,0,128,0C198.7,0,256,57.3,256,128z};
    \fill[white] svg{M86.3,186.2H70.9V79.1h15.4v48.4V186.2z}
                 svg{M108.9,79.1h41.6c39.6,0,57,28.3,57,53.6c0,27.5-21.5,53.6-56.8,53.6h-41.8V79.1z M124.3,172.4h24.5c34.9,0,42.9-26.5,42.9-39.7c0-21.5-13.7-39.7-43.7-39.7h-23.7V172.4z}
                 svg{M88.7,56.8c0,5.5-4.5,10.1-10.1,10.1c-5.6,0-10.1-4.6-10.1-10.1c0-5.6,4.5-10.1,10.1-10.1C84.2,46.7,88.7,51.3,88.7,56.8z};
  }
}

\usepackage[font=footnotesize]{caption}
\usepackage{float}
\usepackage[list=true]{subcaption}
\usepackage{multirow}
\usepackage[export]{adjustbox}

\newcommand\orcid[1]{\href{https://orcid.org/#1}{\mbox{\scalerel*{
\begin{tikzpicture}[yscale=-1,transform shape]
\pic{orcidlogo};
\end{tikzpicture}
}{|}}}}

\usepackage[bookmarks=false]{hyperref}

\def\BibTeX{{\rm B\kern-.05em{\sc i\kern-.025em b}\kern-.08em
    T\kern-.1667em\lower.7ex\hbox{E}\kern-.125emX}}

\makeatletter 
\let\old@ps@headings\ps@headings 
\let\old@ps@IEEEtitlepagestyle\ps@IEEEtitlepagestyle 
\def\confheader#1{%
\def\ps@headings{%
\old@ps@headings%
\def\@oddhead{\strut\hfill#1\hfill\strut}%
\def\@evenhead{\strut\hfill#1\hfill\strut}%
}%
\def\ps@IEEEtitlepagestyle{%
\old@ps@IEEEtitlepagestyle%
\def\@oddhead{\strut\hfill#1\hfill\strut}%
\def\@evenhead{\strut\hfill#1\hfill\strut}%
}%
\ps@headings%
} 
\makeatother 

\begin{document}

\title{Outage Analysis of TAS-NOMA Systems With Multi-Antenna
Users Over $\alpha$-$\mu$ Fading}

\author{\IEEEauthorblockN{Fernando D. Almeida García \orcid{0000-0003-3747-1511}\IEEEauthorrefmark{1}, Maria C. Luna Alvarado \orcid{0000-0001-9352-3668}\IEEEauthorrefmark{2}, Lenin~P.~Jim\'enez~Jim\'enez \orcid{0000-0002-0595-6509
}\IEEEauthorrefmark{2}, \\ Gustavo Fraidenraich \orcid{0000-0002-0517-1496}\IEEEauthorrefmark{2}, Michel D. Yacoub \orcid{0000-0002-5866-5879}\IEEEauthorrefmark{2}, Nathaly V. Orozco Garzón \orcid{0000-0002-5232-7529}\IEEEauthorrefmark{3}, \\ Jos\'e D. Vega-S\'anchez \orcid{0000-0002-3305-1109}\IEEEauthorrefmark{4}, and
Henry R. Carvajal Mora \orcid{0000-0003-0529-8224}\IEEEauthorrefmark{4}}
\IEEEauthorblockA{Wireless and Artificial Intelligence (WAI) Laboratory, National Institute
of Telecommunications (INATEL)\IEEEauthorrefmark{1}\\ School of Electrical and Computer Engineering, University of Campinas (UNICAMP)\IEEEauthorrefmark{2}\\
Faculty of Engineering and Applied Sciences (FICA), Universidad de Las Américas (UDLA)\IEEEauthorrefmark{3}
\\
Colegio de Ciencias e Ingenierías "El Politécnico", Universidad San Francisco de Quito (USFQ)\IEEEauthorrefmark{4}
\\
Email: fernando.almeida@inatel.br\IEEEauthorrefmark{1}}
}



\maketitle

\begin{abstract}
This paper analyzes the outage performance of downlink NOMA systems with transmit antenna selection (TAS) and multi-antenna users over \mbox{$\alpha$-$\mu$} fading. Maximal-ratio combining (MRC) and equal-gain combining (EGC) are considered, with imperfect successive interference cancellation (ipSIC) explicitly modeled. Exact closed-form outage probability (OP) expressions and asymptotic results are derived, offering insights into diversity and coding gains. Simulations validate the analysis, showing that TAS improves the far user’s performance and that MRC outperforms EGC. The results also quantify the loss from ipSIC, and highlight the impact of power allocation on joint OP. The proposed framework serves as a unified tool for evaluating TAS-NOMA systems under generalized fading, providing design insights for B5G/6G networks.
\end{abstract}


\section{Introduction}


Non-orthogonal multiple access (NOMA) has emerged as a key enabling technology for fifth-generation (5G) and beyond (B5G) mobile networks, allowing multiple mobile terminals to communicate simultaneously with a base station (BS) over the same radio resource.
NOMA can be implemented through code-domain schemes, alternative waveforms, or power-domain allocation. 
In the latter, successive interference cancellation (SIC) is employed at the receivers; however, practical limitations result in residual interference, which degrades system performance.

The performance of NOMA systems---characterized in terms of outage probability (OP)---has been extensively studied under diverse transmission schemes (e.g., downlink, uplink, cooperative), fading environments (e.g., Rayleigh, Weibull, Nakagami-$m$, mixture of Gamma (MoG), $\kappa$-$\mu$, $\alpha$-$\mu$, etc.), diversity techniques at the transmitter (e.g., transmit antenna selection (TAS), beamforming (BF)) and receiver (e.g., maximal-ratio combining (MRC), equal-gain combining (EGC)), and under both perfect and imperfect SIC (ipSIC). A summary of representative contributions is provided in Table~\ref{tab: Related Works}, which also indicates the nature of the OP results, i.e., whether exact, approximate, asymptotic, or not derived.

\begin{table*}[!t]
\centering
\caption{Comparison with the state-of-the-art studies}
\label{tab: Related Works}
\begin{tabular}{l c c c c c c c} 
\hline
\vspace{0.05cm}
Paper  & \begin{tabular}[c]{@{}c@{}} NOMA \\ System \end{tabular}   & \begin{tabular}[c]{@{}c@{}} Fading \\ Environment \end{tabular}  & \begin{tabular}[c]{@{}c@{}} Transmission \\ Diversity \end{tabular} & \begin{tabular}[c]{@{}c@{}} Reception \\ Diversity \end{tabular} & \begin{tabular}[c]{@{}c@{}} Imperfect \\ SIC \end{tabular} & OP Analysis   \\ \hline
\vspace{0.05cm} 
\cite{Do20} & Downlink & Rayleigh & TAS & NO & NO & Exact  \\
\vspace{0.05cm} 
\cite{Zhang16,Zhu18,Chen17,Alavi17,Obiedollah19} & Downlink & Rayleigh & BF & NO & NO & NO   \\
\vspace{0.05cm} 
\cite{Mobini22}  & Coop.Downlink &  Rayleigh & TAS & NO & YES & Exact/Asymptotic \\
\vspace{0.05cm} 
\cite{Pei20} & Coop. Downlink & Rayleigh & TAS & NO & YES &  NO \\
\vspace{0.05cm} 
\cite{Jimenez22} & Downlink & Weibull & TAS & EGC & NO &  Exact/Asymptotic  \\
\vspace{0.05cm} 
\cite{Bariah19} & Downlink & Nakagami-$m$ & NO & NO & YES & NO  \\
\vspace{0.05cm} 
\cite{Hou18} & Downlink & Nakagami-$m$ & NO & NO & YES &  Exact/Asymptotic \\
\vspace{0.05cm} 
\cite{Zhang17} & Downlink & Nakagami-$m$ & TAS & MRC & NO &  Exact\\
\vspace{0.05cm} 
\cite{Aldababsa22} & Coop. Downlink & Nakagami-$m$ & TAS & MRC & NO & Exact/Asymptotic \\
\vspace{0.05cm} 
\cite{Le21} & Uplink &  MoG (approx.) & NO & NO & YES  & Approximated/Asymptotic \\
\vspace{0.05cm} 
\cite{Agarwal20} & Uplink/Downlink &  MoG (approx.) & NO & NO & NO & Approximated/Asymptotic \\
\vspace{0.05cm} 
\cite{Neha20}  & Downlink & Double Rayleigh & NO & NO & NO & Exact \\
\vspace{0.05cm} 
\cite{Jaiswal21} & Downlink & Double Nakagami-$m$ & TAS & NO & NO & Exact/Asymptotic \\
\vspace{0.05cm} 
\cite{Yue20},\cite{Shuai22} & Downlink & Shadowed-Rician & TAS & NO & YES  &  Exact/Asymptotic \\
\vspace{0.05cm} 
\cite{ElHalawany20} & Downlink & $\kappa$-$\mu$ Shadowed & NO & NO &  NO & Exact  \\
 \vspace{0.05cm} 
\cite{Sharma19} & Downlink & $\eta$-$\mu$ and $\kappa$-$\mu$ & NO & NO & NO & Exact \\
\vspace{0.05cm} 
\cite{Arias2023NOMA}& Downlink & $\kappa$-$\mu$ & NO & NO &  YES & NO  \\
\vspace{0.05cm} 
\cite{Kumar20} & Downlink & $\alpha$-$\mu$ & NO & NO & NO & NO  \\
\vspace{0.05cm} 
\cite{Li20}  & Coop. Downlink & $\alpha$-$\mu$ & NO & NO & YES &   Exact/Asymptotic \\
\vspace{0.05cm} 
\cite{Arzykulov20}  & Downlink & $\alpha$-$\mu$ & NO &  NO & YES & Exact and asymptotic  \\
\vspace{0.05cm} 
\cite{Alqahtani21} & Downlink & $\alpha$-$\mu$ & NO & NO & NO & Exact  \\
\vspace{0.05cm} 
\textbf{This paper} & \textbf{Downlink} &  $\alpha$-$\mu$ & \textbf{TAS} & \textbf{EGC/MRC} & \textbf{YES} & \textbf{Exact and asymptotic}  \\
\hline
\end{tabular}
\end{table*}

As summarized in Table~\ref{tab: Related Works}, NOMA is often combined with multi-antenna transmission (e.g., TAS) and reception diversity (e.g., MRC, EGC) to improve spectral efficiency and reliability. However, the impact of ipSIC has not been thoroughly addressed in the literature, despite its practical relevance. Moreover, recent works have shown that the $\alpha$-$\mu$ distribution provides an accurate characterization of fading at millimeter-wave frequencies \cite{Marins19,Marins21}, making it an attractive model for future wireless systems. 

To the best of our knowledge, the challenging joint performance evaluation of TAS with MRC/EGC, and ipSIC over $\alpha$-$\mu$ fading channels has not been conducted.
To fill this gap, this work investigates the performance of a multi-antenna downlink NOMA system over $\alpha$-$\mu$ fading with ipSIC. 
Specifically, we analyze the users’ OP  and derive \emph{novel exact and asymptotic expressions}. The functional form of the $\alpha$-$\mu$ model allows for the investigation of both MRC and EGC combining schemes with the same level of mathematical involvement. Table~\ref{tab: Related Works} highlights how our work distinguishes itself from those related in the literature.

In the following, $\text{Pr} \left[  \cdot \right]$ denotes probability; $\mathbb{E} \left[ \cdot \right]$, expectation; $\mathbb{V} \left[ \cdot  \right]$, variance; $\left| \cdot \right|$, absolute value;  $\Gamma(\cdot)$, the gamma function; and $\simeq$, ``asymptotically equal to around zero,'' i.e., ${f(x) \simeq g(x) \iff \lim\limits_{x \to 0} \frac{f(x)}{g(x)} = 1}$.

\section{System Model}
\label{sec: System Model}
Consider a downlink NOMA system with three nodes: a BS equipped with $A$ transmitting antennas and two users equipped with $N$ receiving antennas. The nearest user is denoted as $U_{1}$ and is separated from the BS by a distance of $d_1$, whereas the farthest user is denoted as $U_{2}$ and is separated from the BS by a distance of $d_2$.
Let $h_{k,n,a}$ denote the complex channel coefficient between the $a$-th transmit antenna and the $n$-th receive antenna of the $k$-th user, where $k \in \{1,2\}$.
Also, we assume that (i) each fading envelope (i.e., $|h_{k,n,a}|$) follows an $\alpha$-$\mu$ distribution, and (ii) all the channels experience i.i.d. fading, i.e., $|h_{k,n,a}|$ is independent of $|h_{l,p,b}|$ $\forall (k \neq l,n \neq p ,a \neq b)$. The probability density function (PDF) of $|h_{k,n,a}|$ is given by~\cite{Yacoub07}
\par\nobreak\vspace{-\abovedisplayskip}\small
\begin{align}
    \label{eq: PDF h}
     \mathit{f}_{\left| h_{k,n,a}\right| } (h)=\frac{\left(\alpha  \mu ^{\mu } h^{\alpha  \mu -1}\right) }{\Gamma (\mu ) \hat{h}^{\alpha  \mu }} \exp \left(-\mu  \left(\frac{h}{\hat{h}}\right)^{\alpha }\right),
\end{align}
\normalsize
where $\alpha>0$ is a nonlinearity parameter, $\hat{h}\triangleq \sqrt[\alpha]{\mathbb{E} \left[ |h_{k,n,a}| ^\alpha\right]}$ is the $\alpha$-root mean value, and $\mu>0$ is the inverse of the normalized variance of $|h_{k,n,a}|^\alpha$, i.e., $\mu= \mathbb{E}^2 \left[ |h_{k,n,a}|^\alpha\right] /\mathbb{V} \left[|h_{2,n,a}|^\alpha \right]$.



\subsection{Transmission and Reception Criteria}

Without loss of generality, we assume that the farthest user $U_2$ experiences poorer channel conditions, i.e., $|h_{1,n,a}|^2 > |h_{2,n,a}|^2$.
Accordingly, the BS selects the transmitting antenna that maximizes the signal-to-interference-plus-noise ratio (SINR) of $U_{2}$.\footnote{It is worth noting that the analytical framework provided herein can also be used if one seeks to maximize the SINR of $U_{1}$.}
Once the transmitting antenna has been selected, the BS simultaneously sends the information to both users in the same radio resource (time-frequency), resulting in the overlapping of the signals.
At the receiver side, the multi-antenna users apply either EGC or MRC to combine the received signal replicas. 
Accordingly, the BS selects the transmit antenna that provides the best performance, denoted by $a^*$, based on the following criterion:
\par\nobreak\vspace{-\abovedisplayskip}\small
\begin{align}
\label{eq:arg-max}
    a^*= \underset{1\leq a\leq A}{\text{arg max}} \;\;\left( \sum_{n=1}^{N} |h_{2,n,a}|^{\vartheta_\nu}  \right)^{2/\vartheta_\nu},
\end{align}
\normalsize
where the index $\nu \in \{\text{EGC}, \text{MRC}\}$ specifies the diversity combining scheme, with 
$\vartheta_{\text{EGC}} = 1$ and $\vartheta_{\text{MRC}} = 2$.


On the basis of the NOMA principle, the BS sends the superposed signals to both users. Hence, the received signals at the $n$th antenna of the $k$th user can be written respectively as
\par\nobreak\vspace{-\abovedisplayskip}\small
\begin{align}\label{eq:Y1 Y2}
  y_{k,n}=& \left(\sqrt{\rho P}s_{1} + \sqrt{(1-\rho)P}s_{2}\right)h_{k,n,a^*} + w_{k,n},
\end{align}
\normalsize
where $\rho$ is a power allocation coefficient,\footnote{
As previously stated, the channel condition of $U_1$ is superior to $U_2$, resulting in more power allocated to $U_2$ according to the NOMA principle, i.e., $0 \leq \rho < 0.5$.
} $P$ is the total received power, $s_{k}$ denotes the complex symbol transmitted by the $k$th user belonging to a constellation with unitary mean power, $h_{k,n,a^*}$ is the complex channel coefficient satisfying the criterion in \eqref{eq:arg-max}, and $w_{k,n}$ is the complex additive white Gaussian noise sample with zero mean and variance $N_{0}/T_{s}$ affecting the $n$th branch of the $k$th user. 
In addition, $N_{0}$ is the unilateral noise power spectral density and $T_{s}$ is the symbol duration. 




\subsection{Imperfect Successive Interference Cancellation}

According to the NOMA principle, user $U_1$ decodes its own signal by first canceling the signal of $U_2$ using a successive interference cancellation (SIC) detector.
However, in practice, the signal of $U_2$ cannot be completely eliminated. Thus, by considering the NOMA system model used in \cite{Mahady_2019,Sfredo_2021}, where ipSIC is considered, the resulting signal at the $n$th antenna of $U_1$ can be written as~\cite{Mahady_2019}
\par\nobreak\vspace{-\abovedisplayskip}\small
\begin{align}\label{eq:Y1}
  y_{1,n}=& \left(\sqrt{\rho P}s_{1} + \xi \sqrt{(1-\rho)P}s_{2}\right)h_{1,n,a^*} + w_{1,n},
\end{align}
\normalsize
where $\xi \in \left\{0,1\right\}$ denotes the SIC imperfection level at the receiver of $U_1$. In particular, $\xi = 0$ represents perfect SIC (i.e., no residual interference), and $\xi= 1$ represents ipSIC.
On the other hand, $U_{2}$ decodes its signal directly, that is, without using SIC, because the interference inflicted by $U_{1}$ is small and can be considered as noise \cite{Mahady_2019}.

\subsection{Signal-to-Interference-Plus-Noise Ratio}


At the receiver of $U_1$, the combined signal of $U_1$ 
and $U_2$ is detected. Then, according to the adopted system model, the SINR of $U_1$ is given by $\gamma_{U_1} = \rho/(\xi^2 (1-\rho) + 1/\Phi_{1})$,
where 
\par\nobreak\vspace{-\abovedisplayskip}\small
\begin{align}\label{eq:Phi_1}
    \Phi_{1} =\frac{E_s g_\nu}{N_0} \left[\sum _{n=1}^{N} |h_{1,n,a^*}|^{\vartheta_\nu}\right]^{2/\vartheta_\nu}, 
\end{align}
\normalsize
$E_s= P T_{s}$ is the average energy per symbol, $E_s/N_0$ is the received signal-to-noise ratio (SNR) per symbol, and $\{|h_{1,n,a^{*}}|\}_{n=1}^N$ is the set of i.i.d. $\alpha$-$\mu$ fading envelopes between the transmitting antenna $a^{*}$ and the $n$th receiving antenna of $U_1$ satisfying the criterion in \eqref{eq:arg-max}. Furthermore, $g_{\text{EGC}}=1/N$ and $g_{\text{MRC}}=1$.

At $U_2$, its signal is decoded directly by regarding $U_1$'s signal as interference. Thus, the SINR of $U_2$ is given by $\gamma_{U_2} = (1-\rho)/(\rho + 1/\Phi_{2})$,
where 
\par\nobreak\vspace{-\abovedisplayskip}\small
\begin{align}\label{eq:Phi_2}
    \Phi_{2} = \frac{E_s g_\nu}{N_0} \left[\sum _{n=1}^{N} |h_{2,n,a^*}|^{\vartheta_\nu}\right]^{2/\vartheta_\nu},
\end{align}
\normalsize
and $\{|h_{2,n,a^{*}}|\}_{n=1}^N$ is the set of i.i.d. $\alpha$-$\mu$ fading envelopes between the transmitting antenna $a^{*}$ and the $n$th receiving antenna of $U_2$ satisfying the criterion in \eqref{eq:arg-max}.

\section{Outage Analysis}
\label{sec: Outage Analysis}

In this section, we investigate the performance of a downlink TAS-NOMA system with multi-antenna users over $\alpha$-$\mu$ fading channels. 
To do so, we begin by characterizing $\Phi_1$ and $\Phi_2$.



Recalling that $|h_{1,n,a^{*}}|$ and $|h_{2,n,a^{*}}|$ are independent fading envelopes, and since TAS is applied to maximize the SINR of $U_2$, the choice of the transmitting antenna $a^{*}$ is independent of $U_1$~\cite{Shuai22,Jaiswal21} and (by extension) is also independent of $\Phi_{1}$.
To characterize the PDF of $\Phi_{1}$, in an exact manner, we employ the formulation in~\cite{Garcia2021} for the sum of i.i.d. $\alpha$-$\mu$ random variables. 
By invoking \cite[eq.~(17)]{Garcia2021} and applying successive variable transformations, the PDF of $\Phi_{1}$ is obtained as
\par\nobreak\vspace{-\abovedisplayskip}\small
\begin{align} 
    \label{eq: PDF Phi 1}
    f_{\Phi_{1}}(\phi_1) 
    &= \frac{\vartheta_\nu \beta_\nu}{2 K_\nu}
    \sum_{i=0}^\infty
    \frac{c_i  \left(\frac{\phi_1}{K_\nu}\right)^{\tfrac{\alpha(i+N\mu)}{2}-1}}{\Gamma\!\left(\tfrac{\alpha(i+N\mu)}{\vartheta_\nu}\right)},
\end{align}
\normalsize
where $K_\nu = E_s g_\nu / N_0$, $\beta_\nu=\left(\frac{\alpha\,\mu^\mu}{\Gamma(\mu)\,\vartheta_\nu\,\hat h^{\alpha\mu}}\right)^{\!N}$, and $c_i$ is given in \cite[eq.~(6)]{Garcia2021} under the substitutions 
$\alpha \mapsto \alpha/\vartheta_\nu$ and $\hat h \mapsto \hat h^{\vartheta_\nu}$. 
By integrating \eqref{eq: PDF Phi 1} with respect to $\phi_1$ over the range $0 \leq \phi \leq \phi_1$, the CDF of $\Phi_{1}$ can be expressed as
\par\nobreak\vspace{-\abovedisplayskip}\small
\begin{align}
    \label{eq: CDF Phi 1}
    F_{\Phi_1}(\phi_1) =& \beta_\nu 
    \sum_{i=0}^{\infty}
    \frac{c_i}{\Gamma\!\left(\tfrac{\alpha(i+N\mu)}{\vartheta_\nu}+1\right)}
    \left(\frac{\phi_1}{K_\nu}\right)^{\tfrac{\alpha(i+N\mu)}{2}}.
\end{align}
\normalsize

On the other hand, according to \eqref{eq:arg-max}, the choice of the transmitting antenna $a^{*}$ depends on the statistics of $U_2$. 
In this case, the CDF of $\Phi_{2}$ can be calculated as 
\par\nobreak\vspace{-\abovedisplayskip}\small
\begin{align}
    \label{eq: CDF ph2 step1}
    \nonumber & F_{\Phi_{2}}(\phi_{2}) = \text{Pr} \left[ \frac{E_s}{N N_0} \left[\sum _{n=1}^{N} |h_{2,n,a^*}|^{\vartheta_\nu}\right]^{\frac{2}{\vartheta_\nu}} \leq \phi_{2} \right] \\
    & =  \text{Pr} \left[  \left[\sum _{n=1}^{N} |h_{2,n,1}|^{\vartheta_\nu}\right]^{\frac{2}{\vartheta_\nu}} \leq \delta, \cdots, \left[\sum _{n=1}^{N} |h_{2,n,A}|^{\vartheta_\nu}\right]^{\frac{2}{\vartheta_\nu}} \leq \delta\right],
\end{align}
\normalsize
where $\delta=(\phi_{2} N N_0)/E_s$.
Taking into account that $|h_{2,n,a}|$ and $|h_{2,p,b}|$, $\forall (n \neq p ,a \neq b)$, are i.i.d. fading envelopes, \eqref{eq: CDF ph2 step1} reduces to
\par\nobreak\vspace{-\abovedisplayskip}\small
\begin{align}
    \label{eq: CDF phi2 step2}
    F_{\Phi_{2}}(\phi_{2}) = \left\{ \text{Pr} \left[\frac{E_s g_\nu}{N_0} \left[\sum _{n=1}^{N} |h_{2,n,a}|^{\vartheta_\nu} \right]^{\frac{2}{\vartheta_\nu}} \leq \phi_{2} \right] \right\}^A
\end{align}
\normalsize

Leveraging \cite[eq. (17)]{Garcia2021} and following a procedure analogous to that used for deriving \eqref{eq: CDF Phi 1}, the CDF in \eqref{eq: CDF phi2 step2} can be expressed as
\par\nobreak\vspace{-\abovedisplayskip}\small
\begin{align}
    F_{\Phi_{2}}(\phi_2) 
    &= \beta_\nu^{A}
    \sum_{i=0}^\infty \left[
    \eta_{i,\nu} \left( \frac{\phi_2}{K_\nu}\right)^{\tfrac{\alpha(i+N\mu)}{2}}\right]^A,
    \label{eq: CDF 2 step 1}
\end{align}
\normalsize
\normalsize
where $\eta_{i,\nu}=c_i/  \Gamma\!\left(\tfrac{\alpha(i+N\mu)}{\vartheta_\nu}+1\right)$. 
Now, from \eqref{eq: CDF 2 step 1}, we let $\left[\sum _{i=0}^{\infty } \eta_{i,\nu} \left(\frac{\phi_2}{K_\nu}\right)^{\frac{\alpha i}{2}} \right]^A= \sum _{i=0}^{\infty } \varrho_{i,\nu} \left(\frac{\phi_2}{K_\nu}\right)^{\frac{\alpha i}{2}}$, where the coefficients $\varrho_{i,\nu}$ are determined after differentiating both sides with respect to $\left(\frac{\phi_2}{K_\nu}\right)^{\alpha/2}$, following the approach in~\cite{Garcia23}.
Thus, after extensive algebraic manipulations, the coefficients $\varrho_{i,\nu}$ can be recursively obtained as
\par\nobreak\vspace{-\abovedisplayskip}\small
\begin{align}
    \label{eq: varrho 0}
    \varrho_{0,\nu}=& \frac{  c_0^A}{\Gamma (\frac{N \mu  \alpha }{\vartheta_\nu}+1)^A} \\  \label{eq: varrho i}
    \varrho_{i,\nu} =& \frac{\Gamma (\frac{N \mu  \alpha }{\vartheta_\nu}+1)}{i c_0} \sum_{k=1}^{i} \frac{\left(k A -i+k \right) c_k \varrho_{i-k}}{\Gamma (\frac{k \alpha +N \mu  \alpha }{\vartheta_\nu}+1)} , \ \ \ \ i \geq 0.
\end{align}
\normalsize

Finally,  replacing \eqref{eq: varrho 0} and \eqref{eq: varrho i} into \eqref{eq: CDF 2 step 1}, the CDF of $\Phi_2$ can be obtained as
\par\nobreak\vspace{-\abovedisplayskip}\small
\begin{align}
    \label{eq: CDF 2 final}
    F_{\Phi_{2}}(\phi_{2}) = & \beta_\nu^A \sum _{i=0}^{\infty } \varrho_{i,\nu} \left( \frac{\phi_2}{K_\nu} \right)^{\frac{\alpha}{2} (  i+ \mu A N)}.
\end{align}
\normalsize

The PDF of $\Phi_{2}$ can be obtained by differentiating \eqref{eq: CDF 2 final} with respect to $\phi_{2}$, yielding 
\par\nobreak\vspace{-\abovedisplayskip}\small
\begin{align}
    \label{eq: PDF phi2 final}
    f_{\Phi_{2}}(\phi_{2}) =  \frac{\alpha\,\beta_\nu^{A}}{2K_\nu}
\sum_{i=0}^{\infty}\varrho_{i,\nu}\,\bigl(i+\mu A N\bigr)
\left(\frac{\phi_{2}}{K_\nu}\right)^{\frac{\alpha}{2}(i+\mu A N)-1}.
\end{align}
\normalsize
Due to the versatility of the $\alpha$-$\mu$ distribution, the PDF and CDF in \eqref{eq: PDF phi2 final} and \eqref{eq: CDF 2 final} can also model scenarios in which multi-antenna systems operate under i.i.d. Rayleigh~\cite{Garcia22}, Nakagami-$m$~\cite{Zhang17}, or Weibull~\cite{almeida2021} fading environments.
Now, we proceed to derive the OP of each user, as well as the OP of the overall system.

\subsection{Outage Probability at $U_1$}

The OP of $U_1$ is defined as the probability that the instantaneous SINR falls below a fixed data rate $\tilde{R}_{U_1}$, i.e., $P_{\text{out}}^{U_1}  \triangleq \text{Pr}\left[\gamma_{U_1}\leq\tilde{R}_{U_1} \right ] = F_{\Phi_{1}}\left(\frac{1}{\rho/\tilde{R}_{U_1} - \xi^{2}(1-\rho)}\right)$. Thus, after using \eqref{eq: CDF Phi 1}, we obtain
\par\nobreak\vspace{-\abovedisplayskip}\small
\begin{align}
    \label{eq:Pout_1}
    P_{\text{out}}^{U_1}  = \beta_\nu 
    \sum_{i=0}^{\infty}
    \frac{c_i \left(K_\nu (\rho/\tilde{R}_{U_1} - \xi^{2}(1-\rho)) \right)^{-\tfrac{\alpha(i+N\mu)}{2}} }{\Gamma\!\left(\tfrac{\alpha(i+N\mu)}{\vartheta_\nu}+1\right)}.
\end{align}
\normalsize

The asymptotic OP is useful to address the system performance at high-SNR regime, i.e., when $E_s/N_0 \to \infty$. This is the region where reliable communication systems commonly operate. To obtain the asymptotic OP of $U_1$, we can use the first term of the series in \eqref{eq:Pout_1}, yielding
\par\nobreak\vspace{-\abovedisplayskip}\small
\begin{align}
    \label{eq: asymp OP 1}
    P_{\text{out}}^{U_1} \simeq P_{\text{out},\infty}^{U_1} = \left( O_{\textmd{c},1} \frac{E_s}{N_0}  \right)^{-O_{\textmd{d},1}} , 
\end{align}
\normalsize
where $O_{\textmd{d},1}=\alpha \mu N /2$ is the diversity gain and
\par\nobreak\vspace{-\abovedisplayskip}\small
\begin{align}
    \label{eq: div gain 1}
    O_{\textmd{c},1}= & g_{\nu } \left(\frac{\rho }{\tilde{R}_{U_1}}-\xi ^2 (1-\rho )\right)  \left(\frac{c_0 \beta _{\nu }}{\Gamma \left(\frac{N \alpha  \mu }{\vartheta _{\nu }}+1\right)}\right)^{-\frac{1}{\mathcal{O}_{d,1}}}
\end{align}
\normalsize
is the coding gain for $U_{1}$.

\subsection{Outage Probability at $U_2$}

Similar to user $U_1$, the OP of user $U_2$ can be obtained as $P_{\text{out}}^{U_2} \triangleq   \text{Pr}\left[\gamma_{U_2}\leq \tilde{R}_{U_2} \right ]=F_{\Phi_{2}}\left(\frac{1}{(1-\rho)/\tilde{R}_{U_2}-\rho }\right)$. Thus, using \eqref{eq: PDF phi2 final}, we have 
    \par\nobreak\vspace{-\abovedisplayskip}\small
\begin{align}
    \label{eq:Pout_2}
    P_{\text{out}}^{U_2} =& \beta_\nu^A \sum _{i=0}^{\infty } \varrho_{i,\nu} \left( K_\nu \left( \frac{1-\rho}{\tilde{R}_{U_2}}-\rho \right)\right)^{-\frac{\alpha}{2} (  i+ \mu A N)},
\end{align}
\normalsize
where $\tilde{R}_{U_2}$ is the targeted data rate of $U_2$.
Further, the asymptotic OP of $U_2$ can be found~as
\par\nobreak\vspace{-\abovedisplayskip}\small
\begin{align}
    \label{eq: asymp OP 2}
    P_{\textnormal{out}}^{U_2} \simeq  P_{\text{out},\infty}^{U_2} = \left( O_{\textmd{c},2} \frac{E_s}{N_0}  \right)^{-O_{\textmd{d},2}}, 
\end{align}
\normalsize
where $O_{\textmd{d},2}=\alpha \mu A N/2$ is the diversity gain and
\par\nobreak\vspace{-\abovedisplayskip}\small
\begin{align}
    \label{eq: div gain 1}
   O_{\textmd{c},2}= &  g_{\nu } \left(\frac{1-\rho }{\tilde{R}_{U_2}}-\rho \right) \left(\frac{c_0^A \beta _{\nu }^A}{\Gamma \left(\frac{N \alpha  \mu }{\vartheta _{\nu }}+1\right)^A}\right)^{-\frac{1}{O_{\textmd{d},2}}}
\end{align}
\normalsize
is the coding gain for $U_2$.
Notice from \eqref{eq: asymp OP 1} and \eqref{eq: asymp OP 2} that the diversity gains of both users depend on the fading parameters ($\alpha$ and $\mu$) and the number of receiving antennas ($N$), whereas the diversity gain of $U_2$ takes also into account the number of transmitting antennas ($A$), delivering additional diversity to the user with worst channel conditions.
This additional gain allows us to reduce the allocated power for the second (farthest) user, thus improving the overall system performance.

\subsection{Overall Outage Probability}

For the TAS-NOMA system under consideration, the overall OP is defined as the probability that an outage event occurs at either $U_1$ or $U_2$, and is given by
\par\nobreak\vspace{-\abovedisplayskip}\small
\begin{align}
    \label{eq: OOP def}
    P_{\textnormal{out}}^{\text{overall}} = &  \text{Pr} \left[ \gamma_{U_1} \leq \tilde{R}_{U_1} , \gamma_{U_2} \leq \tilde{R}_{U_2} \right].
\end{align}
\normalsize

From \eqref{eq:Pout_1} and \eqref{eq:Pout_2}, and since $\gamma_{U_1} $ and $\gamma_{U_2}$ are statistically independent, we can finally rewrite \eqref{eq: OOP def} as  
\par\nobreak\vspace{-\abovedisplayskip}\small
\begin{align}
    \label{}
   P_{\textnormal{out}}^{\text{overall}} = & 1 - (1-P_{\text{out}}^{U_1}) (1-P_{\text{out}}^{U_2}).
\end{align}
\normalsize

From \eqref{eq: asymp OP 1} and \eqref{eq: asymp OP 2}, an asymptotic closed-form expression for the overall OP can be derived as
\par\nobreak\vspace{-\abovedisplayskip}\small
\begin{align}
    \label{}
     P_{\textnormal{out}}^{\text{overall}} \simeq P_{\textnormal{out},\infty}^{\text{overall}}   = & 1 - (1-P_{\text{out},\infty}^{U_1}) (1-P_{\text{out},\infty}^{U_2}).
\end{align}
\normalsize

\begin{figure}[t!]
  \centering
  \begin{tabular}[c]{cc}
    \begin{subfigure}[c]{0.23\textwidth}
        \includegraphics[trim={0cm 0cm 0cm 0cm},clip,scale=0.29]{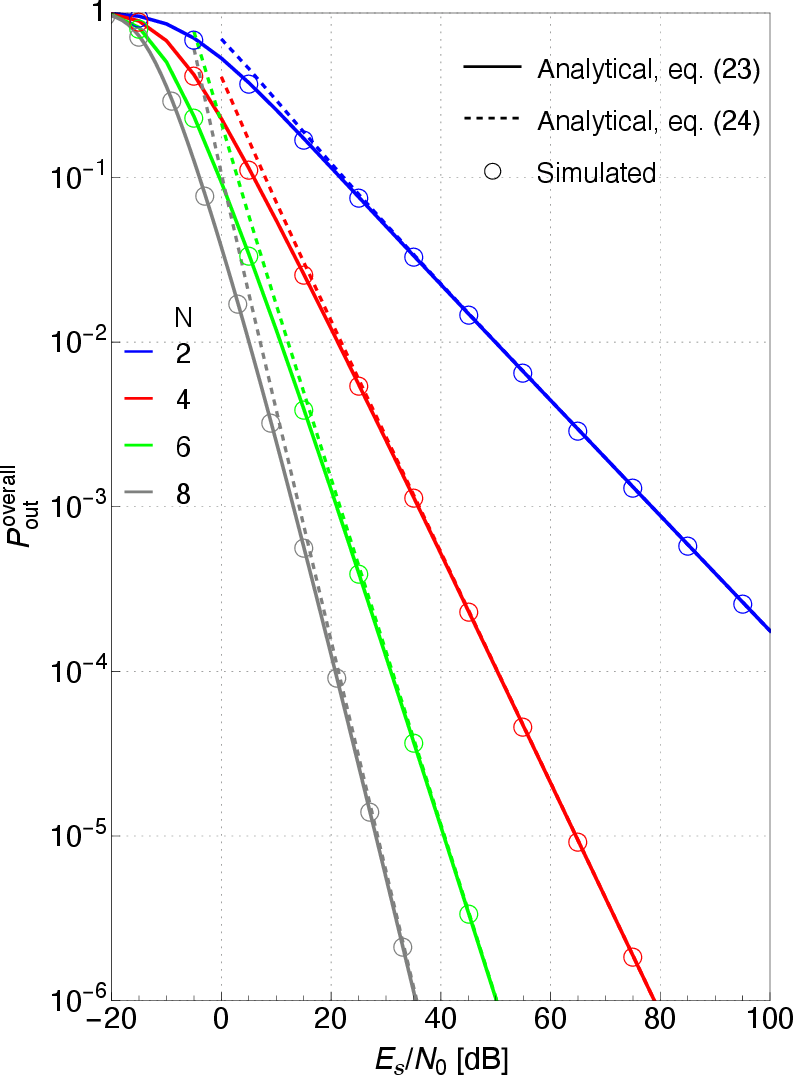}
      \caption{\centering EGC}
    \end{subfigure}
    \hspace{-0.05cm}
    \begin{subfigure}[c]{0.23\textwidth}
     \includegraphics[trim={0.7cm 0cm 0cm 0cm},clip,scale=0.29]{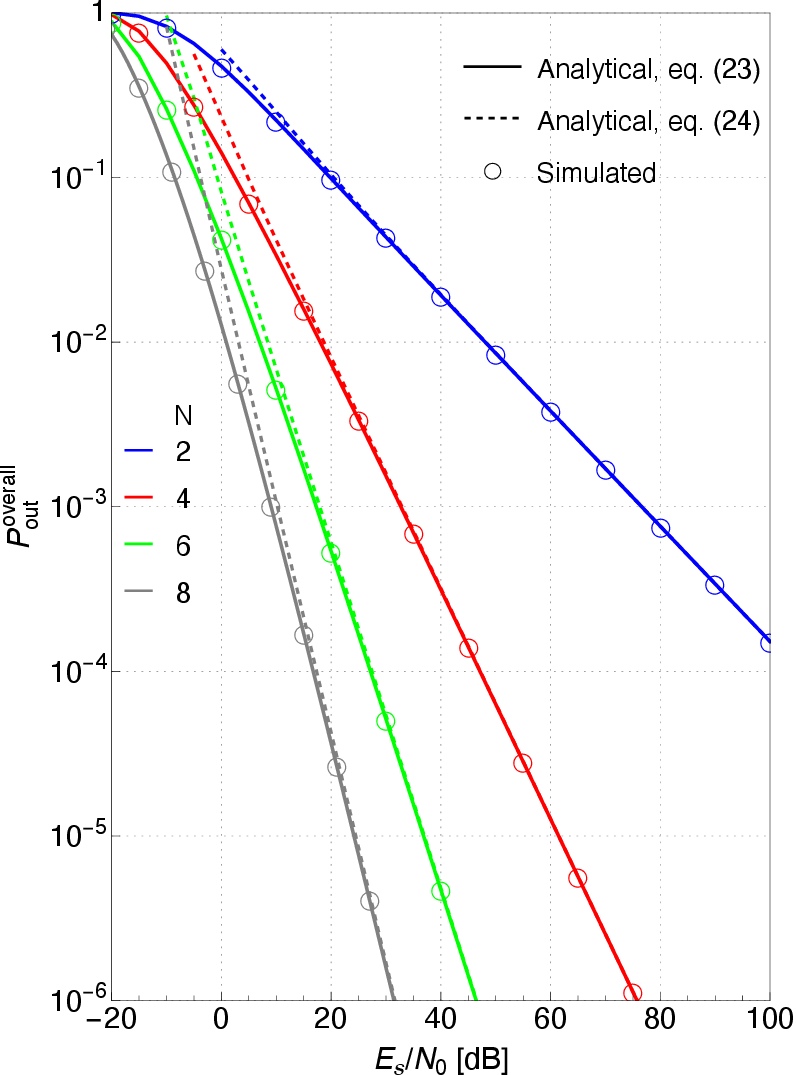}
      \caption{\centering MRC}
      \label{}
    \end{subfigure}
  \end{tabular}
    \caption{Overall OP versus $E_s/N_0$ assuming $\alpha=0.7$, $\mu=0.5$, $A=2$, $\hat{h}=2$, $\xi=1$, $\rho=0.5$, $\tilde{R}_{U_1}=\tilde{R}_{U_2}=0.5$, and different values of $N$.}
  \label{fig: OP N}
\end{figure}
\begin{figure}[t!]
  \centering
  \begin{tabular}[c]{cc}
    \begin{subfigure}[c]{0.23\textwidth}
        \includegraphics[trim={0cm 0cm 0cm 0cm},clip,scale=0.29]{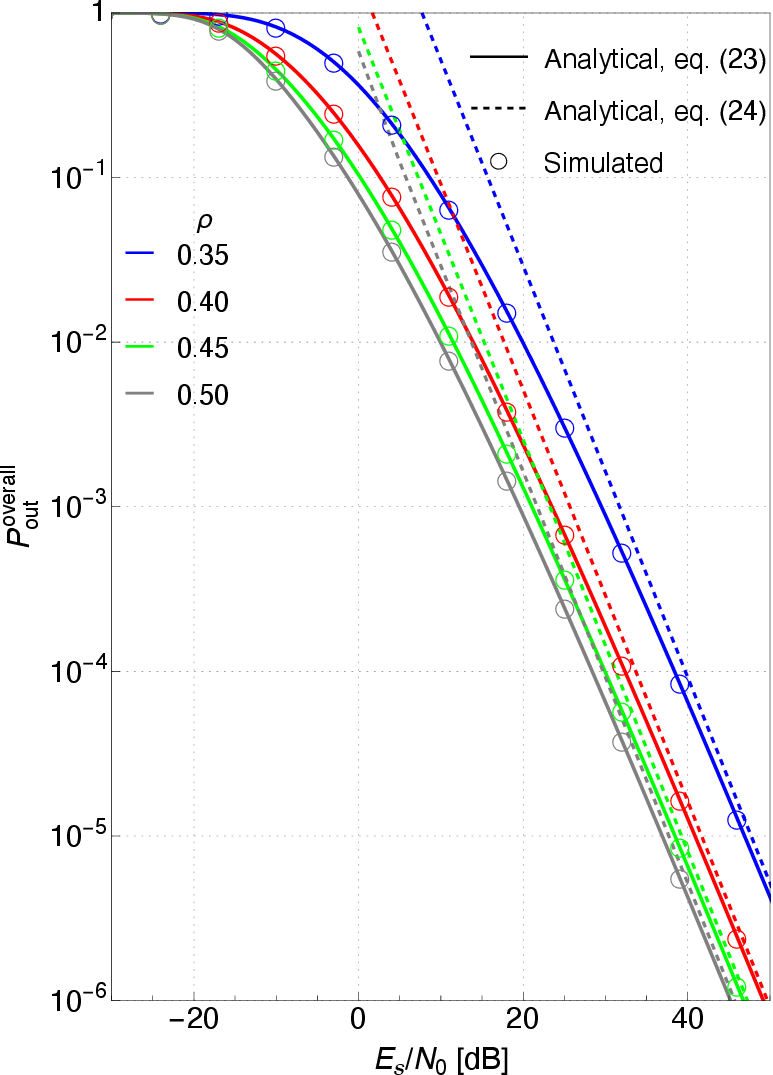}
      \caption{\centering EGC}
    \end{subfigure}
    \hspace{-0.05cm}
    \begin{subfigure}[c]{0.23\textwidth}
     \includegraphics[trim={0.7cm 0.0cm 0cm 0cm},clip,scale=0.29]{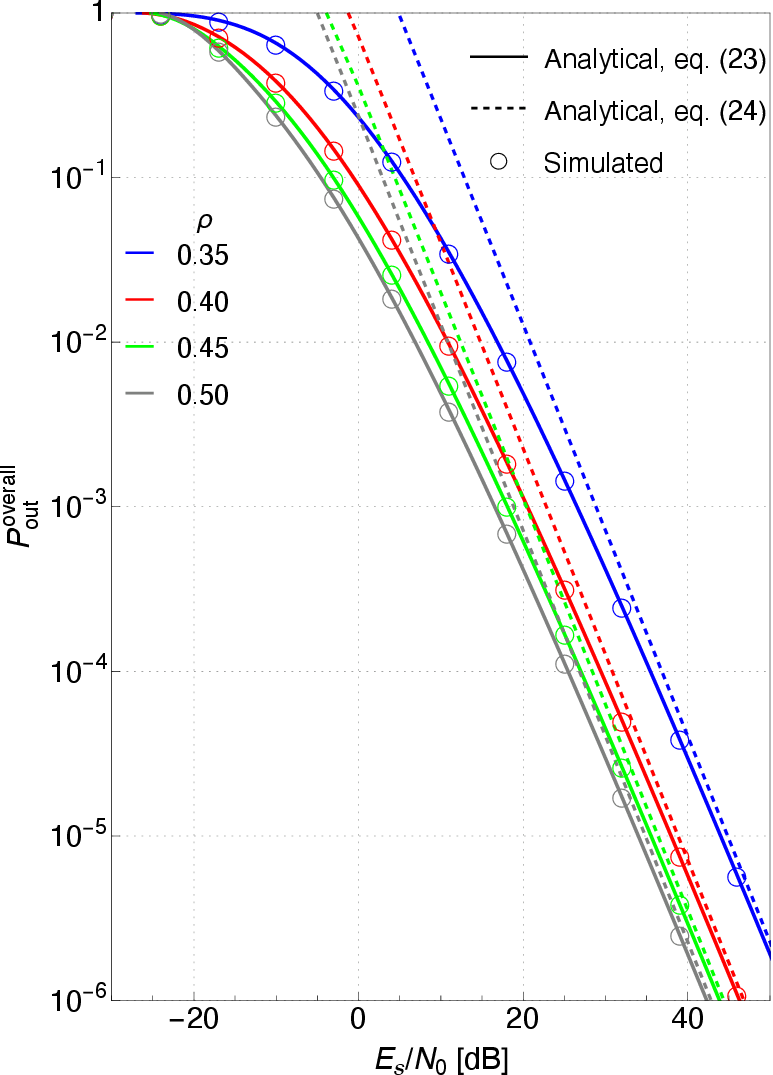}
      \caption{\centering MRC}
      \label{}
    \end{subfigure}
  \end{tabular}
    \caption{Overall OP versus $E_s/N_0$ assuming $\alpha=0.5$, $\mu=1$, $N=5$, $A=3$, $\hat{h}=2$, $\xi=1$, $\tilde{R}_{U_1}=\tilde{R}_{U_2}=0.5$, and various values of $\rho$.}
  \label{fig: OP rho}
\end{figure}

\begin{figure}[t!]
  \centering
  \begin{tabular}[c]{cc}
    \begin{subfigure}[c]{0.23\textwidth}
        \includegraphics[trim={0cm 0cm 0cm 0cm},clip,scale=0.29]{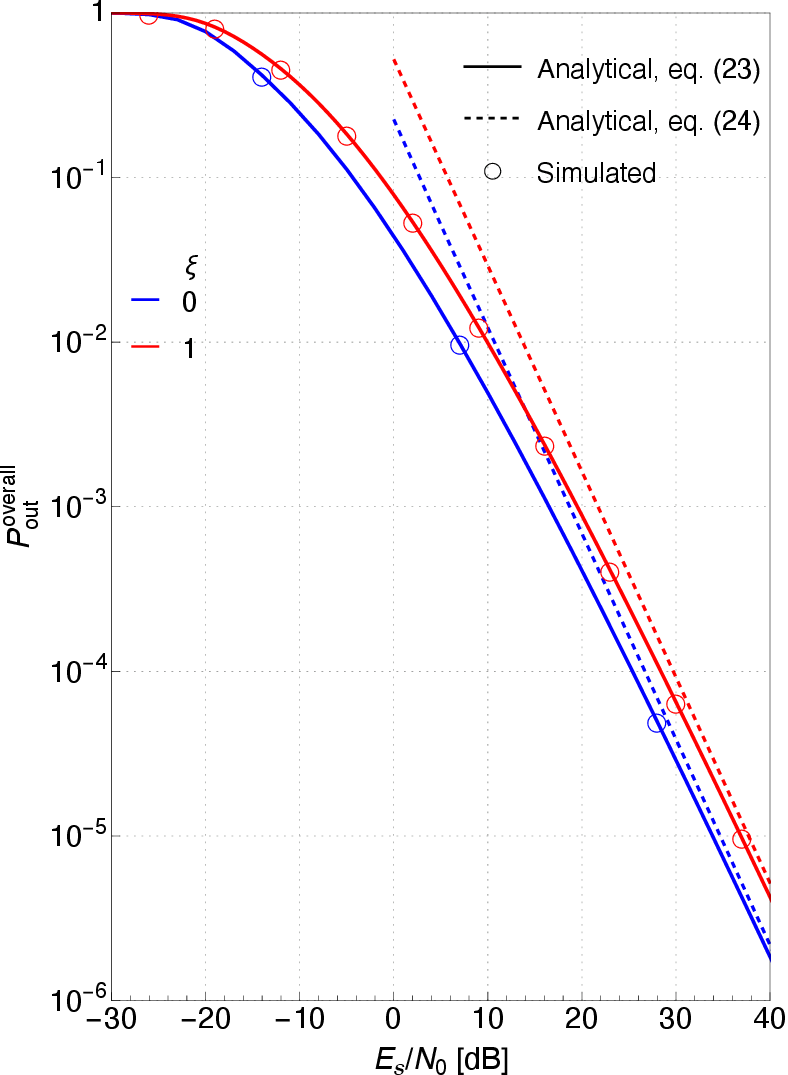}
      \caption{\centering EGC}
    \end{subfigure}
    \hspace{-0.05cm}
    \begin{subfigure}[c]{0.23\textwidth}
     \includegraphics[trim={0.7cm 0cm 0cm 0cm},clip,scale=0.29]{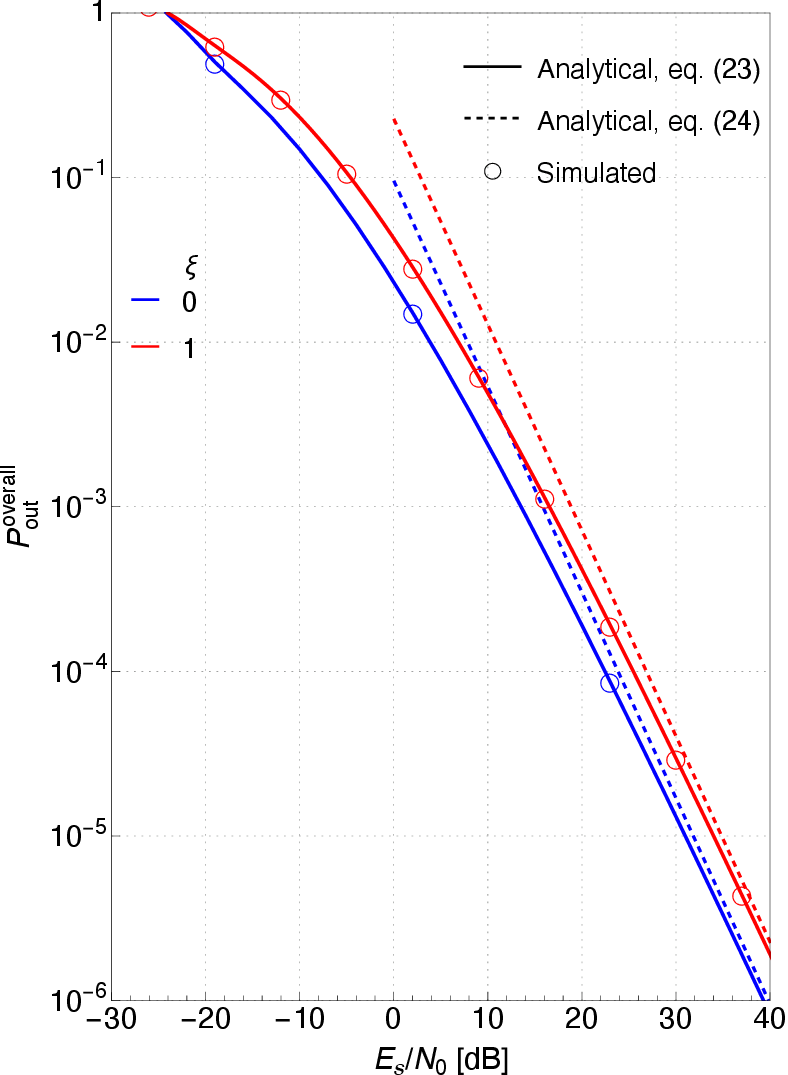}
      \caption{\centering MRC}
      \label{}
    \end{subfigure}
  \end{tabular}
    \caption{Overall OP versus $E_s/N_0$ assuming $\alpha=0.5$, $\mu=1$, $N=5$, $A=7$, $\hat{h}=2$, $\tilde{R}_{U_1}=\tilde{R}_{U_2}=0.5$, and perfect ($\xi=0$) and imperfect ($\xi=1$) SIC.}
  \label{fig: OP xi}
\end{figure}

\section{Sample Results and Discussions}
\label{sec: Numerical Results}

This section validates the analytical findings through Monte Carlo simulations. For all figures, the series expansions of $P_{\text{out}}^{U_1}$ and $P_{\text{out}}^{U_2}$ were computed with 200 terms in their respective series.

Fig. \ref{fig: OP N} shows the overall OP versus $E_s/N_0$ for different values of $N$. As expected, performance improves with the number of receive branches. A pronounced gain is observed when $N$ increases from $2$ to $4$, while the improvements gradually taper off as $N$ increases beyond 4, up to 8, with MRC consistently outperforming EGC, as expected.
Furthermore, the asymptotic curves derived in Section \ref{sec: Outage Analysis} align perfectly with the Monte Carlo results in the high-SNR regime.
Fig. \ref{fig: OP rho} illustrates the overall OP versus $E_s/N_0$ for different power allocation coefficients $\rho$. A trend similar to Fig. \ref{fig: OP N} is observed, where increasing $\rho$ lowers the overall OP
This occurs because allocating more power to $U_1$ reduces its OP, while TAS concurrently improves the performance of $U_2$, yielding a net reduction in the system’s overall OP. Once again, MRC achieves a slight but consistent gain over EGC, and the asymptotic curves align perfectly with the Monte Carlo simulations in the high-SNR region.
Finally, Fig.~\ref{fig: OP xi} presents the overall OP versus $E_s/N_0$ under perfect ($\xi = 0$) and imperfect ($\xi = 1$) SIC. As expected, ipSIC leads to a clear performance degradation, yielding a higher OP for any given $E_s/N_0$. This gap remains evident across both the mid- and high-SNR regimes. Consistently, MRC outperforms EGC, while the asymptotic curves closely match the Monte Carlo simulations in the high-SNR region.

\section{Conclusions}
\label{sec: Conclusions}
This work analyzed the outage performance of TAS-NOMA systems with multi-antenna users over \mbox{$\alpha$-$\mu$} fading channels, considering MRC/EGC combining and the effect of ipSIC. Exact and asymptotic expressions were derived, highlighting diversity and coding gains. Results showed that adding receive antennas improves performance significantly up to four branches, with diminishing returns thereafter. Power allocation strongly impacts joint outage performance, benefiting the near user, while TAS enhances the far user. ipSIC consistently degrades performance, and MRC slightly outperforms EGC. The \mbox{$\alpha$-$\mu$} model enables a unified treatment of MRC and EGC while capturing realistic fading conditions.

\section*{Acknowledgement}
This work was funded by the Brasil 6G Project with support from RNP/MCTI (Grant \mbox{01245.010604/2020-14}), and by the xGMobile Project (Code \mbox{XGM-AFCCT-2025-8-1-1}) with resources from EMBRAPII/MCTI (Grant 052/2023 PPI IoT/Manufatura 4.0) and FAPEMIG Grant \mbox{PPE-00124-23}. The work of M.~C.~L.~Alvarado was supported by the S\~ao Paulo Research~Foundation (FAPESP) under Grant 2023/02578-1. 
The work of L. P. J. Jim\'enez, G. Fraidenraich and M.~D.~Yacoub were supported by the Conselho Nacional de Desenvolvimento Cient\'ifico e Tecnol\'ogico (CNPq) under Grants 141108/2023-1, 302077/2022-7 and  305038/2024-9.
This work was also supported in part by the Universidad San Francisco de Quito (USFQ) through the Poli-Grants Program under Grant 33595. 


\bibliographystyle{IEEEtran}
\bibliography{refs}

\end{document}